\documentstyle[multicol,aps,prb,epsf]{revtex}

\def\lsim{\lower -0.3ex \hbox{$<$} \kern -0.75em \lower 0.7ex \hbox{$\sim$}}
\def\gsim{\lower -0.3ex \hbox{$>$} \kern -0.75em \lower 0.7ex\hbox{$\sim$}}
\newcommand{\Vec}[1]{\mbox{\boldmath$#1$}}

\begin{document}

\draft

\title{Field-Induced SDW and Butterfly Spectrum in Three Dimensions}
\author{M. Koshino, H. Aoki}
\address{Department of Physics, University of Tokyo, Hongo, Tokyo
113-0033, Japan}
\author{T. Osada}
\address{Institute for Solid State Physics, University of Tokyo,
Kashiwa, Chiba 277-8581, Japan}

\date{\today}

\maketitle

\begin{abstract}
Landau's quantization for incompletely nested Fermi surfaces is known to
give rise to magnetic-field-induced spin-density waves(FISDW) in
two-dimensional organic metals. Here we show that three-dimensional(3D)
systems can have 3D-specific series of FISDW phases as energetically
stable states, for which we clarify how and why they appear as the
magnetic field is tilted.  Each phase is characterized by quantized Hall
effect for each of $\sigma_{xy}$ and $\sigma_{zx}$ that reside on a
fractal spectrum like Hofstadter's butterfly.

\end{abstract}

%\newpage

\begin{multicols}{2}
\narrowtext

\section{Introduction}
Rich electronic states arising from nesting of Fermi surfaces 
continue to provide fascination in various classes of materials.  
Organic crystals provide particularly versatile Fermi surfaces, 
and it has indeed been shown\cite{Gorkov} that a curious series of 
spin density wave (SDW) states emerge 
in strong magnetic fields 
in a family of quasi-two-dimensional organic conductors 
(TMTSF)$_2$X (X=PF$_6$ etc), 
called the Bechgaard salt.   The field-induced
spin density wave (FISDW) occurs when 
the nesting of the Fermi surface is incomplete.  
The Landau quantization in the pockets formed 
as a result of an incompletely nesting then causes 
a series of gaps to appear around the main SDW gap\cite{Poil,Maki}.  
Since $E_F$ always lies in the largest Landau gap, an 
integer quantum Hall effect arises. 
When the magnetic field is increased,
successive phase transitions take place 
because the energetically favorable SDW nesting vector 
jumps along the way, which 
results in discontinuous changes of 
the Hall conductivity.  
This has been considered for the TMTSF compound\cite{Ishiguro} 
that happens to have very anisotropic 
transfer energies between molecules with 
$t_x: t_y: t_z \sim 1: 0.1: 0.003$, 
so that the system is almost perfectly two-dimensional(2D).   

So a challenging problem we address here is: 
(i) can we have such Landau-quantization-assisted 
FISDW states in three-dimensional (3D) systems, 
not as a remnant of the 2D FISDW but as 3D-specific, 
energetically favorable states, 
and if so, (ii) how and why 
do the successive phase transitions arise in 3D?  
Lebed'\cite{Lebed} introduced the third direction hopping to FISDW for
the first time, and several authors\cite{Mont,Sun,Hase} 
studied the quantum Hall effect in 
3D FISDW, where Hall conductivities $\sigma_{xy}$ and
$\sigma_{zx}$ are predicted to be quantized respectively.
However, the condition for the emergence of 3D FISDW phase itself
has not been worked out except for a limited case for 
(TMTSF)$_2$X where three dimensionality is very small\cite{Sun}. 
So it has remained to be clarified whether and how FISDW phases 
really do exist in 3D.

This is exactly the purpose of the present paper.   
We consider a possibility of FISDW phases in 3D systems 
in magnetic fields, 
where we shall show that the favorable situation is 
anisotropic 3D systems with an anisotropy such that 
the transfer energies satisfy 
$t_x \gg t_y \sim t_z$ (as contracted with $t_x \gg t_y \gg t_z$ 
in (TMTSF)$_2$X).  
With varied magnitude and orientation of 
the magnetic field $\Vec{B} = (0,B_y,B_z)$, we have 
optimized the SDW nesting vector to show that a series of 3D
FISDW phases do indeed exist, which is best expressed as 
a phase diagram against $(B_y,B_z)$.  The phases comprise 
rich families, where 
they are characterized by quantized Hall conductivities
$\sigma_{xy}$ and $\sigma_{zx}$ as one hallmark of the 
3D-nature.   On the energy axis, the FISDW is seen to reside on a 
fractal energy spectrum like Hofstadter's butterfly\cite{Hofs}, 
which, curiously, also indicates the 3D-specific nature 
of the 3D FISDW.  In fact this can be regarded as 
one realization, through a density-wave formation, of the
butterfly and the quantum Hall effect 
in 3D we have proposed on a general mathematical
basis.\cite{Kosh}   
An intuitive reason why the butterfly spectrum arise in the 3D FISDW is 
discussed in terms of the topology of the incompletely
nested Fermi surface in 3D in the final section.

\section{Formulation for the 3D FISDW}
We consider a simple orthorhombic metal with an energy dispersion
\begin{equation}
\epsilon(\Vec{k}) = -t_x \cos k_x a -t_y \cos k_y b -t_z \cos k_z c,
\end{equation}
where $a,b,c$ are lattice constants and the 
transfer energies are assumed to satisfy $t_x \gg t_y, t_z$ 
(i.e., quasi-1D).
The dispersion along $k_x$ around the Fermi energy can be 
approximated as a linear function $v_F (|k_x| -k_F)$ (with $\hbar = 1$ 
and $\epsilon(\Vec{k})$ measured from $E_F$), 
while the three-dimensionality (warping of the Fermi surface)
can be described by the leading-order expansion in $t_y$ and $t_z$ as
\begin{eqnarray}
\epsilon(\Vec{k}) &=& 
v_F (|k_x| -k_F) + \epsilon_\perp(\Vec{k}_\perp),\\ 
\epsilon_\perp(\Vec{k}_\perp) &=& 
- t_y \cos k_y b -t_z \cos k_z c \nonumber \\
&& - t'_y \cos 2k_y b - t'_z \cos 2k_z c \nonumber \\
&& - t'_{yz} [\cos(k_yb + k_zc) + \cos(k_yb -k_zc)],
\label{dispersion}
\end{eqnarray}
where $\Vec{k}_\perp \equiv (k_y, k_z)$, and
\begin{eqnarray}
t'_y = \alpha t_y^2/t_x, \nonumber \\
t'_z = \alpha t_z^2/t_x, t'_{yz} = 2\alpha t_y t_z /t_x
\label{tprime}
\end{eqnarray}
with
$\alpha = -(\cos k_Fa)/(4\sin^2 k_Fa)$.

Let us apply a magnetic
field $(0,B_y,B_z)$ normal to the conductive axis $x$.
We take the spin quantization axis parallel to $z$.  
We assume that an SDW is the most likely instability as 
in the Bechgaard salts,\cite{super} and 
look at the mean-field equation for the wave function 
with the 3D nesting vector $\Vec{q}= (q_x,q_y,q_z)$ can be written as
\begin{eqnarray}
&&
\left(
\begin{array}{cc}
E - H_{\uparrow}(x) & \Delta(x) \\
\Delta^*(x) & E - H_{\downarrow}(x)
\end{array}
\right)
\left(
\begin{array}{c}
u(x) \\ v(x)
\end{array}
\right)
 = 0 ,
\nonumber \\
&&H_{\uparrow}(x) = 
-i v_F\partial_x + \epsilon_\perp(\Vec{k}_\perp - e\Vec{A}_\perp),
\nonumber \\
&&H_{\downarrow}(x) = 
+i v_F\partial_x + \epsilon_\perp(\Vec{k}_\perp - \Vec{q}_\perp - e\Vec{A}_\perp),
\label{schrodinger}
\end{eqnarray}
where $\Vec{A}_\perp = (B_z x, -B_y x)$ is the vector potential,
and the band energy measured from $- v_F k_F$.
$H_{\uparrow} (H_{\downarrow})$ is the Hamiltonian for
an electron on the right Fermi surface with up-spin 
(or on the left Fermi surface with down-spin), while 
$u (v)$ is the corresponding wave function for 
an up-spin electron on the right Fermi surface 
(down-spin on the left).  
$\Delta(x)$ represents the mean-field electron
interaction, which can be approximately written as a single-mode function
$\Delta(x) \sim \Delta e^{iq_x x}$.  We determine $\Delta$ and $\Vec{q}$ 
self-consistently so as to minimize the free energy
at $T=0$ (i.e., the ground state energy).
The SDW also mixes down-spin states around the right Fermi surface and 
up-spins around left Fermi surface, which defines another order parameter. The
phase difference between the two order parameters specifies the spin order
direction on the $xy$-plane.

If we separate out the $\epsilon_\perp-$dependent phase as
\begin{eqnarray}
u(x) &=& \tilde{u}(x)\exp\left[-\frac{i}{v_F}\int^x_0 
\epsilon_\perp(\Vec{k}_\perp-e\Vec{A}_\perp)
dx'\right], \nonumber \\
v(x) &=& \tilde{v}(x)\exp\left[+\frac{i}{v_F}\int^x_0 
\epsilon_\perp(\Vec{k}_\perp -\Vec{q}_\perp -e\Vec{A}_\perp)
dx'\right], \nonumber \\
\Delta(x) &=& \tilde{\Delta}(x)\exp\bigg(-\frac{i}{v_F}\int^x_0 
[\epsilon_\perp(\Vec{k}_\perp-e\Vec{A}_\perp) \nonumber \\
&& + \epsilon_\perp(\Vec{k}_\perp -\Vec{q}_\perp -e\Vec{A}_\perp)]
dx'\bigg)
\label{substitution}
\end{eqnarray}
Eq.(\ref{schrodinger}) reads
\begin{eqnarray}
\left(
\begin{array}{cc}
E + iv_F\partial_x & \tilde{\Delta}(x) \\
\tilde{\Delta}^*(x) & E - iv_F\partial_x 
\end{array}
\right)
\left(
\begin{array}{c}
\tilde{u}(x) \\ \tilde{v}(x)
\end{array}
\right)
 = 0,
\label{schrodinger2}
\end{eqnarray}
where the effect of the magnetic field is included in the
off-diagonal part, $\tilde{\Delta}$.  When we plug  
Eq.(\ref{dispersion}) into $\tilde{\Delta}$, we obtain
\begin{eqnarray}
  &&\tilde{\Delta}(x) = \Delta e^{iq_x x} \sum_{n_1...n_6}
     J_{n_1}(z_1)J_{n_2}(z_2)\times ... \times J_{n_6}(z_6)
   \nonumber\\
  &&\times
   e^{-i(n_1 + 2n_3 + n_5 + n_6)G_bx - i(n_2 + 2 n_4 + n_5 - n_6)G_c x + 
i\delta}
\label{tilde_delta}
\end{eqnarray}
with
\begin{eqnarray}
&&z_1 = 2t_y/(G_b v_F) \cos (q_y/2), \hspace{0.3cm}
z_2 = 2t_z/(G_c v_F)\cos (q_z/2), \nonumber\\
&&z_3 = t'_y/(G_b v_F) \cos q_y, \hspace{0.3cm}
z_4 = t'_z/(G_c v_F) \cos q_z, \nonumber\\
&&z_5 = t'_{yz}/[(G_b+G_c) v_F] \cos[(q_y+q_z)/2],\nonumber\\
&&z_6 = t'_{yz}/[(G_b-G_c) v_F] \cos[(q_y-q_z)/2],
\label{z}
\end{eqnarray}
where $J_n$ is the Bessel function, 
\[
G_b = eB_zb, G_c = eB_yc, 
\]
and $\delta(q_y,q_z)$ is a phase factor independent of $x$. 
The summation in Eq.(\ref{tilde_delta}) can be rearranged into
\begin{equation}
  \tilde{\Delta}(x) =
   \Delta \sum_{mn} I_{mn} e^{i(q_x - m G_b - n G_c)x + i\delta},
\label{tilde_delta2}
\end{equation}
where $I_{mn}$ is a summation of products of $J_n$'s.  
We can see that the energy gaps of width $|\Delta I_{mn}|$ open at
$k_x = \pm\frac{1}{2}(q_x - mG_b - nG_c)$. 
Since the Fermi energy (at $k_x = \pm k_F$) 
must lie in the largest gap to minimize the energy, we
obtain
\begin{equation}
\frac{1}{2}(q_x - M G_b - N G_c) = k_F,
\label{GMN}
\end{equation}
where $M,N$ are $m,n$ that give the largest $I_{mn}$.
Thus the $x$ component of the SDW nesting vector becomes
$q_x = 2k_F + M G_b + N G_c$.\cite{Lebed}
Here we assume $k_F \gg G_b, G_c$, which is reasonable 
as long as typically $B < 10^4$T.

To be precise, the gaps other than the one at $E_F$ can affect the stability
of the FISDW, but in the weak-coupling regime at $T=0$ 
we can show that the stability of the FISDW phase is 
determined by the width of the gap in which $E_F$ resides.
Suppose $G_b/G_c$ is rational with
\[
G_b = p G, \, \, G_c = q G,
\]
where $p,q$ are mutually prime integers.
Equation (\ref{tilde_delta2}) can then be rewritten as
\begin{equation}
  \tilde{\Delta}(x) =
   \Delta \sum_{l} I_{l} e^{i(q_x - l G)x + i\delta},
\end{equation}
where $I_l$ is the summation of $I_{mn}$ over those $(m,n)$ satisfying
$mp + nq = l$.
The energy spectrum has a gap at 
$k_x = \pm\frac{1}{2}(q_x - l G)$ for each integer $l$.
We consider the situation where 
the gap widths are smaller than the gap intervals. 
We can then express the energy dispersion
along $x$ in the extended zone (shown in Fig. \ref{schem}) as
\begin{eqnarray}
E^\pm(k_x) = \xi^\pm + 
 \sum_l \bigg[{\rm sgn}(\xi^\pm - l \varepsilon)
 \sqrt{(\xi^\pm - l \varepsilon)^2 + |\Delta I_l|^2} \nonumber\\
 - (\xi^\pm - l \varepsilon)\bigg],
\label{E_k}
\end{eqnarray}
where $\xi^\pm (k_x) = \pm \hbar v_F(k_x \mp \frac{1}{2}q_x)$
are the dispersions for $\Delta = 0$ measured from
the gap at $l=0$ for the right ($\xi^+$) and 
left ($\xi^-$) Fermi surfaces with $\varepsilon = \hbar v_F G /2$.
The energy gained by opening the gap in the metallic state is
\begin{equation}
F = \frac{|\Delta|^2}{v_0} + \sum_{\Vec{k},\pm}
\left[ E^\pm(k_x) -\xi^\pm(k_x) \right].
\end{equation}
Here $v_0 (>0)$ is a molecular-field constant, 
and the summation taken over $E_F -\xi_c < E^\pm(k_x) < E_F$, 
where $\xi_c$ is a cutoff.
If we insert Eq.(\ref{E_k}) into this equation, we have
\begin{eqnarray}
F = \frac{|\Delta|^2}{v_0}
- D_0 \frac{|\Delta I_L|^2}{2}
\left( 1+\log\frac{4\xi_c^2}{|\Delta I_L|^2}\right)
\nonumber\\
+ D_0 \sum_{l \ne 0}|\Delta I_{L+l}|^2
\log\left|\frac{l\varepsilon}{\xi_c+l\varepsilon}\right| ,
\label{freeenergy}
\end{eqnarray}
where $L$ is the index of the gap that contains $E_F$,
and $D_0$ is the density of states for $\Delta = 0$ 
which is assumed to be a constant.
From the gap equation, $\partial F / \partial |\Delta|^2 = 0$,
we obtain
\begin{equation}
|\Delta I_L| = 2 \xi_c 
\exp\left(\frac{-1}{|I_L|^2 v_0 D_0} 
+ \sum_{l \ne 0} \left| \frac{I_{L+l}}{I_L} \right|^2
\log\left|\frac{l\varepsilon}{\xi_c+l\varepsilon}\right|\right),
\label{gapeq}
\end{equation}
\begin{equation}
F = -D_0\frac{|\Delta I_L|^2}{2}.
\label{freeenergy2}
\end{equation}
Thus $\Delta$ in general 
depends not only on the width of the gap at $E_F$
($\propto I_L$) but also those of other gaps.
In the weak-coupling limit $v_0 \rightarrow 0$, however, 
$\Delta$ is mainly determined by the factor
$\exp[-1/(|I_L|^2 v_0 D_0)]$.
So larger $I_L$ gives larger $\Delta$ in (\ref{gapeq}),
which gives smaller $F$ in (\ref{freeenergy2}).
Therefore, we only have to maximize $I$ in order to 
minimize the free energy.

\section{Phase Diagram and Hall Conductivity}
We have obtained the phase diagram against $(B_y, B_z)$ 
by maximizing $I_{mn}(q_y, q_z)$ for mesh points on $(B_y, B_z)$ and 
$(q_y, q_z)$ around $(\pi,\pi)$.  
Fig. \ref{phasediagram} shows the result for $t_y = t_z$ (a) and $t_y
> t_z$ (b).  In both cases we do have a series of phases 
that are characterized by $(M,N)$ defined in Eq.(\ref{GMN}).  
An essential finding here is that there are FISDW 
phases {\it specific to 3D}, which exist only when both $t_y$ and
$t_z$ are nonzero.  We can see this by comparing 
Figs. \ref{phasediagram}(a)(b), where the 3D-specific phases (shaded) 
are seen to shrink as $t_z/t_y\rightarrow 0$.  
The 3D-specific phases are classified into several families: 
$(M,N)=(N,-N)$ phases lying along $\theta \equiv {\rm tan}^{-1}(B_y/B_z) =
45^\circ$, and $(-2N,0)$ phases around $(B_y, B_z) \simeq (0.1, 0)$, 
etc, and
their mirror images ($B_y \leftrightarrow B_z$).  
Sun and Maki \cite{Sun} have shown that a small $t_z$ in (TMTSF)$_2$X 
(i.e., $t'_z \propto t_z^2$ neglected) can give rise
to a phase with nonzero $M,N$ just at a particular angle of $\Vec{B}$
(Lebed's angle, corresponding to $45^\circ$ in our model for $b=c$). The
Sun-Maki phase is possibly related to the present 3D phases, although it
does not belong to the $(N,-N)$ family here.

The integers $(M,N)$ have an important physical meaning
--- the Hall conductivity. Following Yakovenko's formulation for
2D\cite{Yakovenko}, Sun and Maki\cite{Sun} have predicted that 
the FISDW phase having $(M,N)$ 
should have Hall conductivities $(\sigma_{xy},
\sigma_{zx}) = \frac{2e^2}{h}(M,N)$ (2: spin factor). 
In our previous paper\cite{Kosh} that demonstrated 
a realization of Hofstadter's butterfly in 
non-interacting 3D systems, 
we have obtained the quantum Hall integers residing 
on the fractal spectrum by making use of Streda's formula 
following 
Halperin-Kohmoto-Wu\cite{Kohm}, where these integers are identified to
be topological invariants assigned to each gap in the butterfly. If we
apply this general argument to the FISDW problem treated here, the result
coincides with Sun-Maki's. 
What is interesting about the FISDW states 
considered here ($t_y \sim t_z$) is that the 
wild variation of $(M,N)$ with the magnetic field accompanies 
a wild variation in the quantum Hall conductivities.

The mathematical origin of the 3D phases can be traced back to 
the basic equations 
above (while we discuss the intuitive reason later).  
For $\theta \rightarrow 45^\circ$, $G_b-G_c$ vanishes and
the argument of one of the Bessel functions, $J_{n_6}(z_6)$, 
diverges. Since $J_n(z)$ has the maximum at $z \sim n$, 
$\tilde{\Delta}(x)$ has a large Fourier component $e^{-in_6(G_b-G_c)x}$
with a nonzero $n_6$.  If we assume other $z$'s are small, $I_{mn}$ has
a maximum at $(m,n) = (n_6,-n_6)$, which corresponds to the $(N,-N)$
phases.  Similarly, $(-2N,0)$ phases correspond to the divergence of
$z_3\propto 1/G_b$.  

Now we come to the stability of the 3D phases.  
When we go from the 3D systems over to 2D ($t_z$ (or $t_y) \rightarrow 0$), 
the 3D phases vanish and we are left with the 2D phases with
$N,0 (0,N)$ that depend only on $B_z$ ($B_y$), as seen 
from Fig. \ref{phasediagram}(b).  These phases are 
known for (TMTSF)$_2$X, while the 3D phases are new.  The
nesting vector $(q_y, q_z)$ is pinned to $(\pi, \pi)$ in the $(N,-N)$
and $(-2N,0)$ phases, while in the 2D $(N,0)$ phases and some of 3D
phases the nesting starts to deviate from $(\pi, \pi)$ with $N$.  
We also notice that the 3D 
phases do not require very large magnetic fields.  In fact, when
$B_y$ or $B_z$ becomes too large the 3D phases give way to 2D ones even
when $t_y \simeq t_z$ as seen in Fig. \ref{phasediagram}(a).  This is
because a large in-plane component of $\Vec{B}$ tends to confine the
electron motion within each layer so that the system becomes 2D-like.

The 3D FISDW phases with larger integers are less
stable since $I_{mn}$ (width of the energy gap) generally
decreases with increasing $m,n$. Hence the FISDW should 
become unstable when the magnetic field 
is too close to $\theta=0, 45^\circ,$ or $90^\circ$,
where the Hall integers diverge. In this region, some metallic phase may 
become stable, or some FISDW with $(q_y, q_z)$ far from $(\pi, \pi)$ may
appear, while we have studied the range $0.9\pi \leq q_x, q_y \leq \pi$
here.

\section{Energy Spectrum}

The second key result in this paper is the quasi-particle spectrum, 
which is plotted
against $B_z/B_y$ in Fig. \ref{btfl}(a). A structure reminiscent of
Hofstader's butterfly are conspicuous around the Fermi energy.  
A closer examination reveals that the whole spectrum, 
consisting of various butterflies pieced together, is much 
more delicately constructed than a single butterfly.  
This is exactly because 
the optimized nesting vector (which jumps from one optimal 
$(M,N)$ to another as $\Vec{B}$ is varied) makes the spectrum 
pieced together in such a way that the Fermi energy 
always lies in the largest gap.
For comparison we display in Fig. \ref{btfl}(b) the energy spectrum 
when the optimization of the nesting vector is neglected 
with a fixed $\Delta$. A zigzag trajectory of the position 
at which the largest gap occurs corresponds to
the gap at $E=0$ in (a).  

We can also trace back the mathematical reason why we have a 
butterfly.  Namely, the quasi-particle equation for the present system
happens to coincide to that for the 3D butterfly in
non-interacting systems previously studied\cite{Kosh},
in that the two periods $G_b, G_c$ (arising from uniform $B_z, B_y$) compete 
with each other, where a difference is that 
the amplitude of the periodicity is here related 
to the order parameter $\Delta$.
So the spectrum plotted against $B_z/B_y$ is in fact expected to have the same
structure as Hofstadter's butterfly revealed in \cite{Kosh}.  An
important distinction from the non-interacting case, however, is that the FISDW
phase adjust itself in such a way that the largest gap in the butterfly
has the Fermi energy in it. 

So, while in the non-interacting case the
butterfly structure is observed only around the bottom (or top) of the
entire band, now we have the butterfly precisely around the Fermi level
{\it by construction}, so the situation should be 
easier to realize experimentally.

\section{Discussions}

{\it Intuitive Picture} --- 
To help understand the butterfly intuitively, we can look at the 
topology of the Fermi surface.  If we first look at the case of the 
3D butterfly in non-interacting systems, 
a typical Fermi surface around the band bottom consists of
nearly parallel planes with a set of holes connecting them as shown in
Fig. \ref{nesting}(a). 
So we end up with, topologically, a coexistence of a bunch of pipes 
$\parallel y$ and another bunch $\parallel z$, 
and this induces a competition between the
Landau quantizations due to $B_y$ and $B_z$, which causes the 3D
butterfly.  If we go back to the present FISDW, we can see 
that the incompletely nested Fermi surface 
has a similar structure after the
SDW gap formation, as typically shown in Fig. \ref{nesting}(b).  
There we display a warped Fermi surface in 3D, where 
the Fermi surface shifted by the nesting vector $\Vec{q}$ is superposed 
to show that how they are interwoven.  
When the SDW gap opens in this incompletely nested Fermi surface in 3D, 
we have a {\it multiply-connected 
Fermi surface} (i.e., a network of pipes) 
reminiscent of Fig. \ref{nesting}(a) as well as 
isolated pockets.  

The situation sharply contrasts with the incompletely nested Fermi surface 
in 2D, where we end up with isolated pipes after the SDW formation. 
Thus the multiply-connected Fermi surface explains how the butterfly-like 
spectrum appears, although, to be more precise, 
there is magnetic breakthrough 
across the pockets and multiply-connected Fermi surface.  
So we expect that the 3D butterfly tends to appear in systems 
having multiply-connected Fermi surfaces.

Figure \ref{nesting}(b) also explains intuitively why SDW gaps are not
formed for magnetic fields having $\theta \sim 0, 45^\circ, 90^\circ$, 
since the semiclassical orbits on
the multiply-connected Fermi surface are open in this case, 
so that the SDW formation is not energetically favorable.  
Mathematically, the divergence of the arguments in Bessel
functions mentioned above is related to the configuration of the Fermi
`pipes'.

{\it Experimental possibilities} --- 
Experimentally, a best region to probe in the phase diagram, 
Fig.\ref{phasediagram} to observe the 3D 
FISDW and the 3D butterfly should be where 
the 3D phase is observed  
for the entire tilting angle ($0<|\theta|<45^\circ$) 
of the magnetic field 
with a fixed $|\Vec{B}|$.  This corresponds to a situation, 
\begin{equation}
t'_y, t'_z \gsim e B b v_F.
\label{crit}
\end{equation}
Why this should be the criterion may be understood as follows.  
The basic equation is written in terms of $z_1 ... z_6$. 
As discussed above, the 3D butterfly is a result of 
a competition between the periods $G_b, G_c$.  
In other words, we need to have $z_3, ..., z_6 \gsim O(1)$, since
$z_3, ...,z_6$ contributes to the Fourier component of $G_b$ or $G_c$
through $J_n(z)$.  We can exclude 
$z_1, z_2$ from our analysis, since they are always small when $(q_y,q_z) 
\simeq (\pi, \pi)$.  
So we end up with the criterion, $t'_y, t'_z \gsim e B b v_F$ from the 
definition of $z_3, ...,z_6$ 
for $b \approx c$.  We do not have to add a condition 
$t'_{yz} (\equiv 2\sqrt{t'_y t'_z}) \gsim e B b v_F$, 
since this condition is already included in the above one.  

We can give a rough idea how we can realize the above condition. 
If we have a material with, say, $t'_y, t'_z \sim 10$K 
(cf. $t'_y \sim 10{\rm K} \gg t'_z$ in (TMTSF)$_2$X) with the values of 
$v_F,b,c$ similar to those in (TMTSF)$_2$X, 
then the butterfly and the peculiar quantum Hall effect should be observed for 
a moderate $B \lsim $ 10T. 
The energy scale of the butterfly will be $t'_y$ or $t'_z$ 
as seen in Fig. \ref{btfl}.  To have a large FISDW gap energy scale, on the 
other hand, larger the $|B|$ the better, since 
for a small magnetic field (for which $z$'s become large) 
$I_{mn}$  has a spreaded distribution 
against $m,n$ and the gaps become smaller.

M.K. would like to acknowledge a Research Fellowship of the Japan 
Society for the Promotion of Science for Young Scientists 
for a financial support.  He also wishes to thank Prof. B.I. Halperin 
and his hospitality at Harvard University where the manuscript is 
completed.

\begin{figure}
%\begin{center}
%  \leavevmode\epsfxsize=80mm \epsfbox{gap_schem.eps}
%\end{center}
\caption{
The structure of the energy spectrum representing 
Eq.(\ref{E_k}) in the text. 
}
\label{schem}
\end{figure}

\begin{figure}
%\begin{center}
%  \leavevmode\epsfxsize=80mm \epsfbox{phasediagram2.eps}
%\end{center}
\caption{
The phase diagram for the FISDW in 3D  at $T=0$ 
in the weak-coupling regime 
is shown against $(B_y,B_z)$ 
for $t_z/t_y=1$ (a) or $0.7$ (b) 
[i.e., $t'_z/t'_y=$1 (a) or 0.49 (b) in eq.(\ref{tprime})].  
The phases are labeled by 
the quantum Hall integers $(M,N) 
[ = (\sigma_{xy},\sigma_{zx})$ in units of $(h/2e^2)$], 
and those having $(q_y,q_z) \ne (\pi,\pi)$ 
are underlined.  
We assume $b=c$, $t_y/t_x = 0.1$ and $\alpha = 0.4$.
The 3D-natured phases are shaded.
}
\label{phasediagram}
\end{figure}

\begin{figure}
%\begin{center}
%  \leavevmode\epsfxsize=80mm \epsfbox{btflFISDW2.eps}
%\end{center}
\caption{
(a) The quasi-particle energy spectrum
against $B_z$ for $t_z/t_y=1$ with $B_y$
fixed to 2.5 (dashed line in Fig. 1).
We assume a coupling constant $v_0D_0 = 0.34$
and the cut-off energy $E_c = 12.5t'_y$.  
Vertical lines indicate boundaries between 
different FISDW phases labeled by $(M,N)$.
(b) Similar spectrum when we do not optimize 
the nesting vector (i.e., $\Vec{q}=(2k_F,\pi,\pi)$) 
with a fixed $\Delta (=0.5 t'_y$ here) for comparison.  
The positions of the gaps having the largest $I_{mn}$ 
are indicated by a solid line.
}
\label{btfl}
\end{figure}

\begin{figure}
%\begin{center}
%  \leavevmode\epsfxsize=80mm \epsfbox{nesting2.eps}
%\end{center}
\caption{
(a)A typical Fermi surface for a non-interacting 
quasi-1D system with $t_x \gg t_y \sim t_z$ and
$E_F \sim t_y, t_z$ from the band bottom.   
(b)A typical Fermi surface (mesh) 
superposed with the nested one (gray) translated by $\Vec{q}$ for 
the 3D FISDW case.  
After the SDW gap opening the Fermi surface consists of 
pockets and a multiply-connected network of pipes.  
Solid lines exemplify open orbits for $\theta = 0, 45^\circ$.}
\label{nesting}
\end{figure}

\end{multicols}
\end{document}